\documentclass[a4paper,11pt]{article}
\usepackage{bgs}

\newcommand{\CC}{\mathcal{C}}

\title{Updated Standard Model Prediction for $K \to \pi \nu \bar{\nu}$ and $\epsilon_K$}

\author[a]{Joachim Brod}
\author[b]{Martin Gorbahn}
\author[c]{Emmanuel Stamou}

\affiliation[a]{Department of Physics, University of Cincinnati, \\ Cincinnati, OH 45221, USA}
\affiliation[b]{Department of Mathematical Sciences, University of Liverpool, \\ Liverpool, L69 7ZL, UK}
\affiliation[c]{Fakult\"at f\"ur Physik, TU Dortmund, \\ D-44221 Dortmund, Germany}

\emailAdd{joachim.brod@uc.edu}
\emailAdd{martin.gorbahn@liverpool.ac.uk}
\emailAdd{emmanuel.stamou@tu-dortmund.de}

\abstract{The rare $K \to \pi \nu \bar{\nu}$ decay modes and the
  parameter $\epsilon_K$ that measures CP violation in Kaon mixing are
  sensitive probes of physics beyond the standard model.  In this
  article we provide the updated standard-model prediction for the
  rare decay modes in detail, and summarise the status of
  standard-model prediction of $\epsilon_K$.  We find $\text{BR}(K^+
  \to \pi^+ \nu \bar \nu) = 7.73(61) \times 10^{-11}$ and
  $\text{BR}(K_L \to \pi^0 \nu \bar \nu) = 2.59(29) \times
  10^{-11}$. The uncertainties are dominated by parametric input.}

\FullConference{%
  BEAUTY2020\\
  21-24 September 2020\\
  Kashiwa, Japan (online)}

\begin{document}
\maketitle

\section{Introduction}

The rare kaon decays $K^+ \to \pi^+ \nu \bar{\nu}$ and $K_L \to \pi^0
\nu \bar{\nu}$ together with indirect CP violation in the neutral kaon
system, parameterised by $\epsilon_K$, are among the cleanest probes
of physics beyond the standard model (SM).  The reason is the
exceptional control over short- and long-distance SM contributions.

The rare kaon decay modes are generated by highly virtual electroweak box and $Z$-penguin diagrams
that can be calculated to high precision in perturbation theory.
Light-quark contributions are strongly suppressed by the GIM mechanism, and the decay matrix elements
can be extracted from precisely measured semi-leptonic kaon decays using approximate isospin symmetry.
The GIM suppression also implies that the decay modes are dominated by internal top-quark exchanges,
which makes these decay modes very sensitive to new sources of flavour violation.

The NA62 collaboration recently reported~\cite{CortinaGil:2021nts, CortinaGil:2020vlo} the measurement of the branching ratio $\mathrm{BR}(K^+ \to \pi^+ \nu \bar{\nu}) = (10.6_{-3.4}^{+3.4}|_{\mathrm{stat}}\pm 0.9_{\mathrm{syst}}) \times 10^{-11}$ that also includes data analysed in previous runs~\cite{CortinaGil:2020vlo} (see Refs.~\cite{Adler:2008zza, Artamonov:2008qb} for the older Brookhaven results).
The best upper bound for the neutral decay mode $\mathrm{BR}(K^+ \to \pi^+ \nu \bar{\nu}) \leq 3.0 \times 10^{-9}$ was obtained
by the JPARC-KOTO~\cite{Ahn:2018mvc,Ahn:2020opg} experiment.

On the other hand, the size of indirect CP violation in the neutral kaon system is experimentally
well-known since many years; it is given by
the parameter $\epsilon_    K = (2.228\pm0.011)e^{i(43.5\pm0.5)^\circ} \times 10^{-3}$~\cite{Zyla:2020zbs}.
A precise SM prediction of $\epsilon_K$ was long impeded by a non-converging perturbation series for the charm-quark contribution; this problem has been solved in Ref.~\cite{Brod:2019rzc}.

In light of the recent measurement it is timely to update the theory prediction of the so called golden rare kaon decays and discuss the recent progress in the prediction of the parameter $\epsilon_K$.
In Sec.~\ref{sec:kpnn} we provide the updated SM prediction for the rare kaon decays. We discuss in detail how the numerical values are obtained, and compare to other recent SM predictions.
The theory status of $\epsilon_K$ is briefly summarised in Sec.~\ref{sec:eK}.

\section{Standard Model Update of $K \to \pi \nu \bar{\nu}$\label{sec:kpnn}}

The effective Hamiltonian relevant for the two rare $K \to \pi \nu \bar{\nu}$ decays is given
by~\cite{Buchalla:1995vs}
\begin{align}
\label{eq:HeffSM}
\mathcal{H}_{\text{eff}} = \frac{4G_F}{\sqrt{2}}
\frac{\alpha}{2\pi\sin^2\theta_w} \sum_{\ell=e,\mu ,\tau} \left(
\lambda_c X^\ell
 + \lambda_t X_t \right) (\bar{s}_L \gamma_{\mu} d_L)
 ( \bar{\nu}_{\ell L} \gamma^{\mu} \nu_{\ell L}) + \text{h.c.}\,.
\end{align}
Here, $G_F$ denotes the Fermi constant, $\alpha$ the electromagnetic
coupling constant, and $\sin\theta_w$ the sine of the weak mixing
angle. The Cabibbo-Kobayashi-Maskawa (CKM) matrix elements are
contained in the parameters $\lambda_i = V_{is}^*
V_{id}^{\phantom{8}}$. The left-handed fermion fields are denoted by
$f_L \equiv (1-\gamma_5)/2 f$. The loop functions $X^\ell$ and $X_t$
are discussed below.


The branching ratio of the charged mode is given by
\begin{equation}\label{eq:BR:ch}
  \text{Br} \left(K^+\to\pi^+\nu\bar{\nu}(\gamma)\right) = \kappa_+
  (1+\Delta_{\text{EM}})
  \Bigg[\left(\frac{\text{Im}\lambda_t}{\lambda^5} X_t\right)^2 +
  \left(\frac{\text{Re}\lambda_c}{\lambda} \left(P_c + \delta P_{c,u}
    \right) + \frac{\text{Re}\lambda_t}{\lambda^5} X_t\right)^2
  \Bigg].
\end{equation}
Here, $X_t$ is a function of $x_t = m_t(\mu_t)^2/M_W^2$ and has been
calculated including next-to-leading order QCD~\cite{Misiak:1999yg,
  Buchalla:1998ba} and electroweak~\cite{Brod:2010hi}
corrections.
Hence, the top-quark mass, $m_t$, and the $W$-boson mass,
$M_W$, depend on the QCD and electroweak renormalization schemes. In
fact, $M_W$ is not a primary input and has to be calculated as a
function of the $Z$-boson mass, $M_Z$, the Higgs-boson mass, $M_h$, and
the strong and the electromagnetic coupling constants $\alpha_s$ and
$\alpha$, respectively (see Ref.~\cite{Awramik:2003rn} for more details).
The $\overline{\text{MS}}$ scheme is the natural choice regarding QCD.
We obtained the numerical value $m_t(m_t) = 162.83(67)\,$GeV from the
top-quark pole mass (see Tab.~\ref{tab:param}) by converting it to
QCD-$\overline{\text{MS}}$ at three-loop accuracy, using
RunDec~\cite{Chetyrkin:2000yt}. The electroweak corrections are
minimised if all masses are renormalized on-shell with respect to the
electroweak interactions~\cite{Brod:2010hi}; we adopt this scheme for
our numerics. We obtain our numerical value for $X_t$ by calculating
a mean value of the QCD contribution, $X_t^{\text{QCD},\text{avg}}$,
by varying $\mu_t \in [60,320]$ GeV in the expression
\begin{equation}
  X_t^{\text{QCD}}  = X_t^{(0)}(x_t(\mu_t)) + \frac{\alpha_s(\mu_t)}{4\pi} X_t^{(1)}(x_t(\mu_t))\,,
\end{equation}
and taking the average of the smallest and largest value of $X_t^{\text{QCD}}$.
Here, $X_t^{(0)}$ and $X_t^{(1)}$ denote the leading-order (LO) and next-to-leading-order
(NLO) QCD contributions to $X_t$, respectively.
Electroweak corrections are then taken into account by including the
fit function $r_X$, which is valid for electroweak-onshell masses and
given in Ref.~\cite{Brod:2010hi}. In total
\begin{equation}
  X_t =  X_t^{\text{QCD},\text{avg}} + [r_X(m_t(m_t))-1] \, X_t^{(0)}(m_t(m_t)) \,.
\end{equation}

The theory uncertainty associated with the QCD corrections is given by
the difference of the central value $X_t^{\text{QCD},\text{avg}}$
and
the minimal / maximal value in the $\mu_t$ interval. The uncertainty
associated to the electroweak corrections is $\pm 0.00134 \times
X_t$ \cite{Brod:2010hi}. In total, we find
\begin{equation}
  X_t
  = 1.462 \pm 0.017_{\text{QCD}} \pm 0.002_{\text{EW}}
  \,.
\end{equation}

The parameter $P_c = \lambda^{-4} (\tfrac{2}{3}X^e+\tfrac{1}{3}X^{\tau})$
comprises the charm-quark contribution and has been calculated at
next-to-next-to-leading order (NNLO) in QCD~\cite{Buras:2006gb} and at
NLO in the electroweak interactions~\cite{Brod:2008ss}. It is a
function of $x_c = m_c(\mu_c)^2/M_W^2$ which, upon inclusion of the
electroweak corrections, is defined as $x_c = \sqrt{2} \sin^2\theta_w
G_F m_c^2(\mu_c)/(\pi\alpha)$. A fit formula for $P_c$ and its theory
uncertainty, including the NLO electroweak and NNLO QCD correction in
dependence on the strong coupling and the charm-quark mass, has been
presented in Ref.~\cite{Brod:2008ss}. With the current PDG input, we
find
\begin{equation}
  P_c = \bigg(\frac{0.2255}{\lambda}\bigg)^4 \times (0.3604 \pm 0.0087) \,.
\end{equation}
The effects of dimension-eight operators at the charm threshold, as
well as additional long-distance contributions arising from up- and
charm-quarks have been estimated in Ref.~\cite{Isidori:2005xm},
leading to the correction $\delta P_{c,u} = 0.04(2)$. These effects
can be computed using lattice QCD in the future~\cite{Isidori:2005tv}
(see Ref.~\cite{Bai:2018hqu} for preliminary results).

The hadronic matrix element is contained in the parameter
\begin{equation}
  \kappa_+ = \bigg(\frac{0.231}{\sin^2\theta_w}\bigg)^2
             \bigg(\frac{\alpha(M_Z)}{127.9^{-1}}\bigg)^2
             \bigg(\frac{\lambda}{0.225}\bigg)^8
             \times 0.5173(25) \times 10^{-10} \,,
\end{equation}
extracted from $K_{\ell 3}$ decay including higher-order chiral
corrections~\cite{Mescia:2007kn}. The NLO QED
corrections~\cite{Mescia:2007kn} are parameterised by
$\Delta_{\text{EM}} = -0.003$ in Eq.~\eqref{eq:BR:ch}.

The remaining parametric input is contained in the CKM factors
$\lambda_t$ and $\lambda_c$, defined above. We expand these parameters
in $\lambda$, including the quadratic corrections, and find
\begin{equation}
  \begin{split}
  \text{Im}\lambda_t &= A^2\bar{\eta}\lambda^5 + \frac{1}{2} A^2 \bar{\eta}\lambda^7 + {\cal O}(\lambda^9)\,,\\
  \text{Re}\lambda_t &=
       A^2\lambda^5(\bar{\rho} - 1)
     + \frac{1}{2} A^2\lambda^7(2\bar{\eta}^2 + 2\bar{\rho}^2 - 3\bar{\rho} + 1)
     + {\cal O}(\lambda^9)\,,\\
     \text{Re}\lambda_c &= - \lambda+\frac{1}{2}\lambda^3  + {\cal O}(\lambda^5) \,.
  \end{split}
\end{equation}
As the rare kaon decay modes do not enter the standard global CKM fit,
we use the values obtained from the global fit as input
parameters. All input values are taken from
pdgLive~\cite{Zyla:2020zbs} and are collected here in
Tab.~\ref{tab:param}. We find the following prediction for the charged
mode in the SM,
\begin{equation}
  \text{BR}(K^+ \to \pi^+ \nu \bar \nu) = 7.73(16)(25)(54) \times 10^{-11} \,.
\end{equation}
The errors in parentheses correspond to the remaining short-distance,
long-distance, and parametric uncertainties, with all contributions
added in quadrature. In more detail, the leading contributions to the
uncertainty are
\begin{equation}
\begin{split}
  10^{11} \times \text{BR}(K^+ \to \pi^+ \nu \bar \nu)
  & = 7.73
          \pm 0.12_{X_t^{\text{QCD}}}
          \pm 0.01_{X_t^{\text{EW}}}
          \pm 0.11_{P_c}
          \pm 0.24_{\delta P_{cu}}
          \pm 0.04_{\kappa_+}\\
 & \quad
          \pm 0.13_{\lambda}
          \pm 0.46_{A}
          \pm 0.18_{\bar\rho}
          \pm 0.03_{\bar\eta}
          \pm 0.05_{m_t}
          \pm 0.15_{m_c}
          \pm 0.05_{\alpha_s}
\,.
\end{split}
\end{equation}

\begin{table}
\centering
\begin{tabular}{|ll|ll|}
\hline
$m_t^{\rm Pole}$[GeV] & $172.4\pm0.7$       & $\alpha^{5\text{fl}}_s(M_Z)$ & $0.1179\pm 0.0010$\\
$M_h$[GeV]            & $125.10\pm0.14$     & $\alpha^{-1}(M_Z)$           & $127.952\pm0.009$\\
$M_Z$[GeV]            & $91.1876\pm 0.0021$ & $\Delta\alpha^{\text{had}}$  & $0.02766\pm 0.00007$\\
$m_c$[GeV]            & $1.27\pm 0.02$      & $s_{w,\text{ND}}^{2}(M_Z)$   & $0.23141\pm 0.00004$\\
                      &                     & $|\epsilon_K|$               & $(2.228\pm 0.011) \times 10^{-3}$\\[1em]
$\bar\rho$  &   $0.141\pm 0.017$   & $\lambda$   & $0.22650  \pm 0.00048$\\
$\bar\eta$  &   $0.357\pm 0.011$   & $A$         & $0.790\pm 0.017$\\
\hline
  \end{tabular}
  \caption{Parametric input used for our SM prediction of the $K\to\pi\nu\bar\nu$ branching rations; all values are
    taken from pdgLive~\cite{Zyla:2020zbs}. Using this input, we find
    $m_t(m_t) = 162.83\,$GeV and
    $M_W = 80.36\,$GeV
    (see text for details).
  \label{tab:param}}
\end{table}


The branching ratio of the neutral mode is computed from
\begin{equation}
  \text{Br}\left(K_L\to\pi^0\nu\bar\nu\right)=
  \kappa_L r_{\epsilon_K}
  \left(\frac{\text{Im}\lambda_t}{\lambda^5}X_t \right)^2\,,
  \label{eq:brkL}
\end{equation}
it depends to a good approximation only on the top-quark function $X_t$
discussed above. The hadronic matrix element is contained in the
parameter
\begin{equation}
  \kappa_L = \bigg(\frac{0.231}{\sin^2\theta_w}\bigg)^2
             \bigg(\frac{\alpha(M_Z)}{127.9^{-1}}\bigg)^2
             \bigg(\frac{\lambda}{0.225}\bigg)^8
             \times 2.231(13) \times 10^{-10} \,,
\end{equation}
again extracted from $K_{\ell 3}$ decay including higher-order chiral
corrections~\cite{Mescia:2007kn}.

At the current level of accuracy, also the small contribution of
indirect CP violation~\cite{Buchalla:1996fp} should be included. It is
taken into account in Eq.~\eqref{eq:brkL} by the factor
\begin{equation}
  r_{\epsilon_K} \equiv 1-\sqrt{2}|\epsilon_K|\frac{1+P_c/A^2X_t-\rho}{\eta}\,,
\end{equation}
where $A$, $\rho = \bar{\rho}/(1-\lambda^2/2+\dots)$, and $\eta = \bar{\eta}/(1-\lambda^2/2+\dots)$ are
Wolfenstein parameters. The loop function
$X_t$ and all remaining parametric input has been discussed above in
the context of the charged mode.
Our SM prediction for the neutral mode then reads
\begin{equation}
  \text{BR}(K_L \to \pi^0 \nu \bar \nu) = 2.59(6)(2)(28) \times 10^{-11} \,.
\end{equation}
Again, the errors in parentheses correspond to the remaining
short-distance, long-distance, and parametric uncertainties, with all
contributions added in quadrature. In more detail, the leading
contributions to the uncertainty are
\begin{equation}
\begin{split}
  10^{11} \times \text{BR}(K_L \to \pi^0 \nu \bar \nu)
 & = 2.59 \pm 0.06_{X_t^{\text{QCD}}}
          \pm 0.01_{X_t^{\text{EW}}}
          \pm 0.02_{\kappa_L}\\
 & \quad  \pm 0.16_{\bar\eta}
          \pm 0.22_{A}
          \pm 0.04_{\lambda}
          \pm 0.02_{m_t}
\,.
\end{split}
\end{equation}


Next we discuss the differences between the theory prediction of
Ref.~\cite{Buras:2015qea} and our analysis.
The largest discrepancy arises from the choice of numerical values for the CKM parameters.
The choice of $|V_{ub}|$, $|V_{cb}|$ and $\gamma$ in Ref.~\cite{Buras:2015qea}
implies central values for $\bar{\rho}_{\text{\scriptsize \cite{Buras:2015qea}}}=0.119$ and
$\bar{\eta}_{\text{\scriptsize \cite{Buras:2015qea}}}=0.394$
that deviate from the PDG values for
the SM CKM-fit by roughly $1$- and $3$-$\sigma$ for $\bar\rho$ and $\bar\eta$, respectively.
The difference in the numerical value for $X_t$ has a milder impact on the
branching ratios.
Here, the most recent, improved measurement of $M_t$ and $\alpha_s$ and the corresponding
change of their central values results in a reduction of $X_t$
compared to older determinations.
The error due to unknown higher-order QCD corrections is estimated using
a different range for varying the matching scale, $\mu_t$, which also
implies a slightly lower (0.7\%) $X_t$ value in our determination.
The fact that only approximate results for the electroweak corrections
have been included in Ref.~\cite{Buras:2015qea} is negligible.

\section{Status of $\epsilon_K$\label{sec:eK}}

In this section, we give a brief overview of the recent progress in the
SM prediction for indirect CP violation in the neutral kaon system.
We define the parameter $\epsilon_K \equiv e^{i\phi_\epsilon} \sin\phi_\epsilon \frac{1}{2} \arg ( -M_{12}/\Gamma_{12} )$,
where $\phi_\epsilon = \arctan(2\Delta M_K/\Delta\Gamma_K)$, with $\Delta M_K$
and $\Delta\Gamma_K$ the mass and lifetime differences of the weak
eigenstates $K_L$ and $K_S$.
$M_{12}$ and $\Gamma_{12}$ are the Hermitian and anti-Hermitian parts of
the Hamiltonian that determines the time evolution of the neutral kaon system.
The evaluation of the matrix element
$M_{12} = - \langle K^0 | \mathcal{L}^{\Delta \mathrm{S} = 2}_{f=3}| \bar K^0 \rangle / (2\Delta M_K)$ can,
at leading order in the operator-product expansion, be factorised
into short- and long-distance contributions that can be
calculated in perturbation theory and on the lattice, respectively.
Note that the ratio $M_{12}/\Gamma_{12}$, and hence $\epsilon_K$, does not depend
on the phase convention of the CKM matrix.
To make this apparent, we factor out $1/(\lambda_u^{*})^2$ and $1/(\lambda_u^{*})$ from the $|\Delta \mathrm{S}=2|$ and $|\Delta \mathrm{S}=1|$ effective Lagrangians, respectively, and use CKM unitarity to
express the effective three-flavor $|\Delta S\!=\! 2|$ Lagrangian in terms of
the minimal number of independent CKM parameters.
The resulting Lagrangian with {\em manifest CKM unitarity}~\cite{Brod:2019rzc},
\begin{equation}\label{eq:HS2:inv}
	\mathcal{L}_{f=3}^{\Delta \mathrm{S} = 2} = - \frac{G_{\mathrm{F}}^2M_W^2}{4\pi^2}
  \frac{1}{(\lambda_u^*)^2}  {Q}_{\mathrm{S}2}
  \Big\{ f_{1} \,\CC_{1}(\mu)
    + i J \left[ f_{2} \,\CC_{2}(\mu)
                + f_{3}\,\CC_{3}(\mu) \right] \Big\}
     + \text{h.c.} + \ldots\,,
\end{equation}
is written in terms of three real Wilson coefficients $\CC_{i}(\mu)$, $i=1,2,3$,
four real, independent, rephasing-invariant parameters $J$, $f_{1}$, $f_{2}$, and
$f_{3}$ comprising the relevant CKM matrix elements, and the operator
$Q_{\mathrm{S}2} = (\overline{s}_L \gamma_{\mu} d_L) \otimes (\overline{s}_L \gamma^{\mu}d_L)$.
Explicitly, we have $J = \mathrm{Im}(V_{us} V_{cb} V^*_{ub} V^*_{cs})$ and $f_{1} = |\lambda_u|^4 + \dots$, where the ellipsis denotes real terms that are suppressed by powers of $\lambda$.
In this form it is evident that $\CC_1$, the Wilson coefficient relevant for $\Delta M_K$, does not contribute to $\epsilon_K$.
In the PDG phase convention, the choice of $f_{2} = 2 \mathrm{Re}(\lambda_t \lambda_u^*)$ and $f_{3} = |\lambda_u|^2$ results in the effective Lagrangian
\begin{equation}\label{eq:LS2:final}
\begin{split}
	\mathcal{L}^{\Delta S=2}_{f=3} = - \frac{G_F^2 M_W^2}{4 \pi^2}
    \big[ \lambda_u^2 C_{{S}2}^{uu}(\mu) + \lambda_t^2 C_{{S}2}^{tt}
    (\mu)+ \lambda_u \lambda_t C_{{S}2}^{ut}(\mu) \big] Q_{{S}2}
      + \textrm{h.c.} + \dots \,,
\end{split}
\end{equation}
with the Wilson coefficients $C_{\mathrm{S}2}^{uu} \equiv \CC_1$, $C_{\mathrm{S}2}^{tt} \equiv \CC_2$, and $C_{\mathrm{S}2}^{ut} \equiv \CC_3$.
Only the coefficients $C_{\mathrm{S}2}^{ut}$ and $C_{\mathrm{S}2}^{tt}$ are relevant for the prediction of $\epsilon_K$.
This choice of $f_2$ and $f_3$ results in their respective Wilson coefficients being free of the
low-energy contributions of $\CC_1$ and hence they can be calculated with high precision
in renormalization-group improved perturbation theory.
The higher-order corrections can be conveniently parameterised by the formally
scale-independent parameters $\eta_{tt}$ and $\eta_{ut}$ that encode the
higher-order QCD corrections to the LO Inami--Lim functions
$S_{tt}(x_t)$ and $S_{ut}(x_c, x_t)$ (see Refs.~\cite{Inami:1980fz, Brod:2019rzc}).
Their values are $\eta_{tt} = 0.55(2)$ and $\eta_{ut} = 0.402(5)$~\cite{Brod:2019rzc}.

The SM prediction for the absolute value of $\epsilon_K$ is then obtained via the phenomenological formula~\cite{Buchalla:1995vs, Buras:2008nn, Buras:2010pza}
\begin{equation}
    |\epsilon_K|
=  \kappa_\epsilon C_\epsilon \widehat{B}_K
|V_{cb}|^2 \lambda^2 \bar \eta \\
 \times \Big(|V_{cb}|^2(1-\bar\rho)
\eta_{tt} S_{tt}(x_t) - \eta_{ut} S_{ut}(x_c, x_t) \Big)\,.
\end{equation}
Here, the kaon bag parameter comprising the hadronic matrix element of
the local $\Delta S = 2$ operators is given by $\widehat{B}_K =
0.7625(97)$~\cite{Aoki:2019cca}. The phenomenological parameter
$\kappa_\epsilon = 0.94(2)$~\cite{Buras:2010pza} comprises
long-distance contributions beyond the lowest order in the operator-product
expansions, which are not included in $B_K$, see also Ref.~\cite{Cata:2003mn}
for the calculation of dimension-eight operator matrix elements. The remaining
parametric input is collected in the factor $C_\epsilon =
(G_{F}^2 F_K^2 M_{K^0} M_W^2)/(6\sqrt{2}\pi^2\Delta M_K)$.

The parameter $\epsilon_K$ is one of the main ingredients of the
global CKM fit. Hence, we do not use the CKM parameters extracted from
the global fit for its SM prediction, and instead directly use the PDG
values of $\lambda$, $V_{cb}$, $\sin2\beta$, as well as lattice input
for the ratio of $B$-meson decay constants and bag factors
$\xi_s$~\cite{Aoki:2019cca}, see Ref.~\cite{Brod:2019rzc} for
details. We find
\begin{equation}
  |\epsilon_K| = \big(
		       2.161
		\pm 0.153_{\text{param.}}
		\pm 0.076_\text{non-pert.}
		\pm 0.065_\text{pert.}
		\big)\times 10^{-3} \,.
\end{equation}

\section{Conclusions}

We have presented updated SM predictions of the branching ratios for the
rare kaon decay modes, finding $\text{BR}(K^+ \to \pi^+ \nu \bar \nu)
= 7.73(61) \times 10^{-11}$ and $\text{BR}(K_L \to \pi^0 \nu \bar \nu)
= 2.59(29) \times 10^{-11}$ (all uncertainties have been added in
quadrature), and briefly discussed the current theory status of the
perturbative contribution to $\epsilon_K$.

The perturbative uncertainties in the SM predictions can be further
reduced by calculating the three-loop QCD corrections for the top-quark
contributions to both the rare decays and $\epsilon_K$, as well as
electroweak corrections to the $|\Delta S = 2|$ effective
Lagrangian. These projects are work in progress by the authors.
The improved SM theory prediction together with the current experimental
progress will increase the sensitivity to physics beyond the SM.
Given this sensitivity, it is interesting to note that contributions to the
rare kaon decays and $\epsilon_K$ in a wide class of renormalizable theories
beyond the SM have been presented in a general form in
Refs.~\cite{Brod:2019bro, Bishara:2021buy}.

\subsubsection*{Acknowledgements}

MG would like to thank the organisers of BEAUTY 2020 for their invitation.
JB, MG and ES acknowledge support by the DOE grant DE-SC0011784, the UK STFC under Consolidated Grant
ST/T000988/1, and the COST Action CA16201 PARTICLEFACE.

\bibliographystyle{JHEP}
\bibliography{references}

\end{document}